\documentclass[sigconf]{acmart}
\AtBeginDocument{%
  }

\copyrightyear{2026}
\acmYear{2026}
\setcopyright{cc}
\setcctype{by}
\acmConference[ICSE-SEIP '26]{2026 IEEE/ACM 48th International Conference on Software Engineering}{April 12--18, 2026}{Rio de Janeiro, Brazil}
\acmBooktitle{2026 IEEE/ACM 48th International Conference on Software Engineering (ICSE-SEIP '26), April 12--18, 2026, Rio de Janeiro, Brazil}
\acmPrice{}
\acmDOI{10.1145/3786583.3786853}
\acmISBN{979-8-4007-2426-8/2026/04}

\usepackage{enumitem}
\usepackage{multirow}
\usepackage{makecell}
\usepackage{subfigure}
\usepackage{balance}
\usepackage{listings}
\usepackage{color}
\usepackage{algorithm}
\usepackage{algorithmic}

\usepackage{amssymb}

\begin{document}
	
\author{Lingzhe Zhang}
\affiliation{%
	\institution{Peking University}
	\city{Beijing}
	\country{China}}
\orcid{0009-0005-9500-4489}
\email{zhang.lingzhe@stu.pku.edu.cn}

\author{Tong Jia$^{\ast}$}
\thanks{*Corresponding author}
\affiliation{%
	\institution{Peking University}
	\city{Beijing}
	\country{China}}
\orcid{0000-0002-5946-9829}
\email{jia.tong@pku.edu.cn}

\author{Yunpeng Zhai}
\affiliation{%
	\institution{Alibaba Group}
	\country{China}}
\orcid{0000-0002-3344-4543}
\email{zhaiyunpeng.zyp@alibaba-inc.com}

\author{Leyi Pan}
\affiliation{%
	\institution{Tsinghua University}
	\country{China}}
\orcid{0009-0008-0859-2203}
\email{panly24@mails.tsinghua.edu.cn}

\author{Chiming Duan}
\affiliation{%
	\institution{Peking University}
	\city{Beijing}
	\country{China}}
\orcid{0009-0008-4422-6323}
\email{duanchiming@stu.pku.edu.cn}

\author{Minghua He}
\affiliation{%
	\institution{Peking University}
	\city{Beijing}
	\country{China}}
\orcid{0000-0003-4439-9810}
\email{hemh2120@stu.pku.edu.cn}

\author{Mengxi Jia}
\affiliation{%
	\institution{Institute of Artificial Intelligence}
	\country{China}}
\orcid{0000-0002-0979-9803}
\email{jiamx1@chinatelecom.cn}

\author{Ying Li$^{\ast}$}
\affiliation{%
	\institution{Peking University}
	\city{Beijing}
	\country{China}}
\orcid{0000-0002-6278-2357}
\email{li.ying@pku.edu.cn}

\renewcommand{\shortauthors}{Lingzhe Zhang, et al.}

\title[Agentic Memory Enhanced Recursive Reasoning for Root Cause Localization in Microservices]{Agentic Memory Enhanced Recursive Reasoning for \\ Root Cause Localization in Microservices}

\begin{abstract}
  As contemporary microservice systems become increasingly popular and complex—often comprising hundreds or even thousands of fine-grained, interdependent subsystems—they are experiencing more frequent failures. Ensuring system reliability thus demands accurate root cause localization. While many traditional graph-based and deep learning approaches have been explored for this task, they often rely heavily on pre-defined schemas that struggle to adapt to evolving operational contexts. Consequently, a number of LLM-based methods have recently been proposed. However, these methods still face two major limitations: shallow, symptom-centric reasoning that undermines accuracy, and a lack of cross-alert reuse that leads to redundant reasoning and high latency. In this paper, we conduct a comprehensive study of how Site Reliability Engineers (SREs) localize the root causes of failures, drawing insights from professionals across multiple organizations. Our investigation reveals that expert root cause analysis exhibits three key characteristics: recursiveness, multi-dimensional expansion, and cross-modal reasoning. Motivated by these findings, we introduce AMER-RCL, an agentic memory enhanced recursive reasoning framework for root cause localization in microservices. AMER-RCL employs the Recursive Reasoning RCL engine, a multi-agent framework that performs recursive reasoning on each alert to progressively refine candidate causes, while Agentic Memory incrementally accumulates and reuses reasoning from prior alerts within a time window to reduce redundant exploration and lower inference latency. Experimental results demonstrate that AMER-RCL consistently outperforms state-of-the-art methods in both localization accuracy and inference efficiency.
\end{abstract}



\maketitle

\section{Introduction}

Modern microservices have become increasingly complex due to dynamic interactions and evolving runtime environments~\cite{zhou2018fault, zhang2025survey}. These systems often consist of thousands of fine-grained, interdependent subsystems, where issues in any one component can easily lead to performance problems at the top level~\cite{mendoncca2019developing, waseem2021design, zhang2024towards, zhang2024time, zhang2025microremed, kang2022separation}. Therefore, to ensure system reliability, it is crucial to localize the root cause of these issues in a timely manner~\cite{zhang2024multivariate, zhang2024reducing}.

Localizing root causes in microservice systems is inherently challenging due to the intricate dependencies between subsystems~\cite{wang2023interdependent, zhang2024failure, yu2024survey, sun2025interpretable, zhu2024hemirca, xie2024microservice, wang2024kgroot, zhang2025log, he2025walk, he2025united, duan2025logaction}. Each request typically traverses complex invocation chains that involve services, instances, hosts, and diverse interactions such as service calls and database queries. The dynamic nature of these interactions, coupled with system heterogeneity, makes root cause localization highly non-trivial. To address this, prior research has largely relied on graph-based and deep learning-based approaches. Graph-based methods capture inter-service dependencies through causality or dependency graphs, often enhanced with statistical or machine learning techniques~\cite{lin2018microscope, li2022causal, yu2021microrank, yu2023tracerank, zhang2022crisp, zhang2024trace, lin2024root}. These methods provide interpretable reasoning paths but rely on rigid graph structures, which hinder adaptability across heterogeneous platforms and evolving environments. In contrast, deep learning-based methods automatically learn temporal, spatial, and structural dependencies from metrics and traces~\cite{gan2019seer, yang2022micromilts, cai2021modelcoder, ding2023tracediag, wang2023root, chen2021trace, ren2023grace}. However, these approaches often function as black boxes, making their predictions difficult to interpret and limiting the transferability of learned models across deployments. Overall, despite their effectiveness, both paradigms suffer from common limitations, including lack of cross-platform generality and insufficient adaptability to evolving system states.
 
Fortunately, large language models (LLMs) have shown strong potential to address these limitations, and a growing body of work has begun to explore their application to root cause analysis~\cite{zhang2024mabc, wang2024rcagent, pei2025flow, li2025coca, wang2025tamo, ren2025multi, roy2024exploring, shi2024enhancing, xie2024cloud, han2024potential, zhang2025adaptive, zhang2025thinkfl, zhang2025scalalog, zhang2025agentfm, zhang2024automated, shan2024face}. Existing studies can be broadly categorized into three lines of research: (i) early explorations, which highlight the feasibility and potential of LLMs for root cause analysis~\cite{sarda2024leveraging, roy2024exploring, shi2024enhancing, xie2024cloud, han2024potential, zhang2025agentfm}; (ii) systematic solutions, which employ multi-agent architectures to structure the reasoning process~\cite{zhang2024mabc, wang2024rcagent, pei2025flow, wang2025tamo, ren2025multi, zhang2025adaptive, zhang2025thinkfl}; and (iii) knowledge-enhanced approaches, which integrate techniques such as retrieval-augmented generation (RAG) to ground LLMs in domain knowledge~\cite{zhang2025scalalog, zhang2024automated, sarda2024leveraging, shan2024face, li2025coca}. Despite these promising directions, the practical deployment of LLM-based root cause localization in real-world microservice systems continues to face several critical challenges.

\begin{itemize}[leftmargin=*]
	\item \textbf{Shallow Symptom-Centric Reasoning Limits Accuracy.} 
	Root cause localization requires not only identifying anomalous symptoms but also distinguishing true causes from downstream effects in complex service dependencies. Existing LLM-based approaches often adopt generic reasoning frameworks such as ReAct or Chain-of-Thought (CoT)~\cite{zhang2024mabc, wang2024rcagent, pei2025flow, li2025coca, wang2025tamo, ren2025multi}. While some incorporate task-specific heuristics, their reasoning processes often remain shallow and symptom-centric, failing to systematically differentiate between primary and secondary anomalies. Consequently, these approaches often conflate symptoms with true causes, yielding suboptimal accuracy in complex or cascading fault scenarios.
	\item \textbf{Lack of Cross-Alert Reuse Causes Redundant Reasoning and High Latency.} 
	Many failures in microservice systems are systemic or stochastic (e.g., resource contention, intermittent network issues, or cascading failures) and can only be reliably detected by analyzing aggregate patterns across multiple alerts~\cite{lin2018microscope, li2022causal, lin2024root, yu2021microrank, yu2023tracerank, zhang2022crisp, zhang2024trace}. Existing LLM-based approaches typically analyze each alert independently, triggering full multi-agent reasoning for every alert~\cite{zhang2024mabc, wang2024rcagent, pei2025flow, li2025coca, wang2025tamo, ren2025multi}. For example, mABC~\cite{zhang2024mabc} performs multi-agent reasoning separately for each alert. In large anomalous windows containing hundreds or thousands of alerts, many traces, metrics, or log entries overlap, yet these methods do not reuse prior analyses or share intermediate reasoning results. This leads to substantial redundant computation, high resource consumption, and increased inference latency.
\end{itemize}

Recognizing this gap, we first conducted an empirical study to investigate how SREs localize root causes, providing insights into how they perform in-depth analysis to address the first challenge. This study reveals three key characteristics of manual root cause analysis: (i) \textbf{recursiveness}, where identification of a deeper-level root cause prompts SREs to iteratively refine their analysis by examining lower-layer manifestations; (ii) \textbf{multi-dimensional expansion}, in which the search is broadened both vertically (across system layers) and horizontally (across related components and services); and (iii) \textbf{cross-modal reasoning}, whereby a candidate root cause suggested by trace data is further validated through correlated fluctuations in relevant metrics and logs.

Building on these insights, we introduce \textbf{AMER-RCL}, an \textbf{A}gentic \textbf{M}emory \textbf{E}nhanced \textbf{R}ecursive reasoning framework for \textbf{R}oot \textbf{C}ause \textbf{L}ocalization in microservices. To address the first challenge, AMER-RCL employs the \textbf{Recursive Reasoning RCL} engine, a multi-agent framework that performs recursive reasoning on each alert to generate a ranked set of candidate root causes. To tackle the second challenge, AMER-RCL processes multiple alerts within a time window of a microservice system. Each alert is first transformed into a graph representation via a Causal Graph Extraction process, capturing the anomalous request path and its contextual dependencies. Leveraging these graphs, \textbf{Agentic Memory} incrementally records prior reasoning outcomes and intermediate deliberations, enabling the system to skip redundant analysis for highly similar alerts while focusing recursive exploration on significant divergences, thereby improving both accuracy and efficiency.

We conduct experiments on the AIOPS 2022, Train-Ticket, and FAMOS-Mall datasets to evaluate AMER-RCL. The results show that, in terms of accuracy, AMER-RCL outperforms both other LLM-based and non-LLM-based root cause localization methods, surpassing the second-best approach by approximately 16\%.In terms of inference efficiency for batch alert analysis, AMER-RCL achieves more than 3.5× speedup over existing LLM-based methods. In summary, the key contributions of this work are as follows:

\begin{itemize}[leftmargin=*]
	\item We conduct a comprehensive study on how SREs localize the root cause. Our findings reveal that human root cause analysis exhibits three key characteristics: recursiveness, multi-dimensional expansion, and cross-modal reasoning.
	\item Inspired by these findings, we propose AMER-RCL, an agentic memory enhanced recursive reasoning framework for root cause localization in microservices.
	\item We evaluate AMER-RCL on three datasets to demonstrate its effectiveness. Experimental results show that AMER-RCL surpasses state-of-the-art methods in both root cause localization accuracy and inference efficiency.
\end{itemize}

\section{Background}

In this section, we present the essential background of this paper, including the formal definition of the root cause localization problem, an overview of traditional root cause localization methods, and an introduction to distributed tracing as the primary data source for root cause localization.

\subsection{Root Cause Localization}

Failure diagnosis in distributed systems is generally divided into two categories: failure category classification and root cause localization. The former determines the failure type, such as CPU or memory anomalies, while the latter pinpoints the specific node, service, or pod responsible for the issue. Root cause localization is particularly critical for minimizing downtime and ensuring system stability.

Traditional root cause localization methods typically analyze a collection of alerts within a predefined time window. These methods construct a dependency graph where nodes represent system components, and edges capture causal relationships inferred from traces or system logs. Graph-based algorithms, such as PageRank, are then applied to rank components based on their likelihood of being the root cause.

\begin{equation}
	C^* = C^* = \arg\max_{C \in \mathcal{C}} s(C, G, M)
	\label{eq: rcl}
\end{equation}

Formally, given a set of anomalous requests $R=\{r_1, r_2, ..., r_n\}$, a component graph $G=(C, E)$ is constructed, where $C$ is the set of components, and $E$ represents their dependencies. The root cause component $C^*$ is identified as Equation~\ref{eq: rcl}, where $s(C, G, M)$ represents the computed score incorporating graph topology and observed anomalies.

\subsection{Distributed Tracing}

To support fine-grained fault diagnosis in software systems, distributed tracing has been widely adopted in industrial environments, becoming an integral part of modern software infrastructures~\cite{wang2022characterizing, yang2022capturing, shen2023network}. A distributed trace provides a detailed execution record of a request as it propagates through the system, capturing timing, dependencies, and performance characteristics. Each trace consists of multiple spans, which document individual operations along the request's execution path.

\vspace{-0.8em}
\begin{figure}[htbp]
	\centering
	\includegraphics[width=1\linewidth]{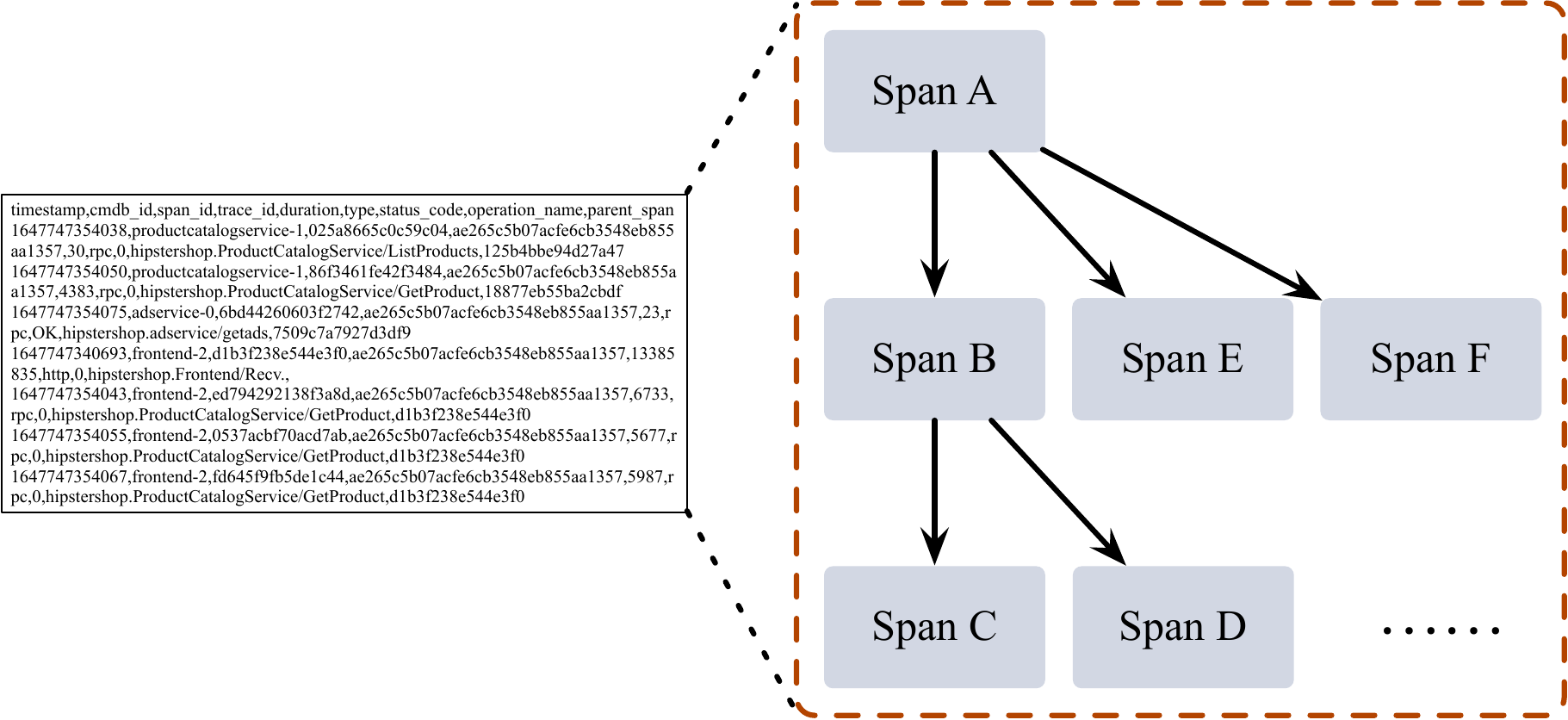}
	\vspace{-1.2em}
	\caption{Example of a Trace}
	\label{fig: tracing}
	\vspace{-0.8em}
\end{figure}

As illustrated in Figure~\ref{fig: tracing}, a complete trace represents the end-to-end journey of a request, detailing every intermediate operation and its corresponding latency. Each span records crucial information such as the service name, operations, timestamps, and causal relationships between operations. By analyzing the timing and dependencies of spans, distributed tracing enables precise performance monitoring and facilitates anomaly detection in large-scale distributed systems. This structured representation is particularly valuable for diagnosing latency issues, identifying service bottlenecks, and uncovering failure propagation patterns across microservices.

\section{Empirical Study}

In this section, we conduct a comprehensive study on how SREs localize the root cause. Our methodology was designed to combine both expert-driven modeling and practitioner validation.

First, one researcher from each of three organizations—Peking University, Tsinghua University, and Alibaba Group—jointly collaborated to analyze representative cases and formulate an initial characterization of the root cause localization process. 

Second, to refine and validate this framework, we conducted semi-structured interviews with a total of 22 professional developers and SREs drawn from Peking University, Tsinghua University, Alibaba Group, and the Institute of Artificial Intelligence at China Telecom. During these interviews, participants were asked to examine the initial framework, reflect on their own operational experiences, and provide corrections or enhancements. The collected feedback was then consolidated into the final version of the framework, ensuring both accuracy and broad applicability.

Through this two-stage process, we arrived at a validated characterization of manual root cause localization. To systematically analyze the process, we organize our findings around the following research questions:

\begin{itemize}[leftmargin=*]
	\item \textbf{RQ1:} How do SREs initially narrow down the search space for the root cause?
	\item \textbf{RQ2:} How do they systematically expand the search scope to encompass all potential root causes?
	\item \textbf{RQ3:} How do they differentiate the actual root cause from other potential but non-causal anomalies?
\end{itemize}

\subsection{Recursiveness}

To narrow down the search space for the root cause, SREs begin with the problematic request and examine its trace data to identify the invoked services and executed operations. If anomalies are detected in these services, they further investigate the downstream services they call, recursively refining the search until the true root cause is isolated.

\vspace{-0.8em}
\begin{figure}[htbp]
	\centering
	\includegraphics[width=1\linewidth]{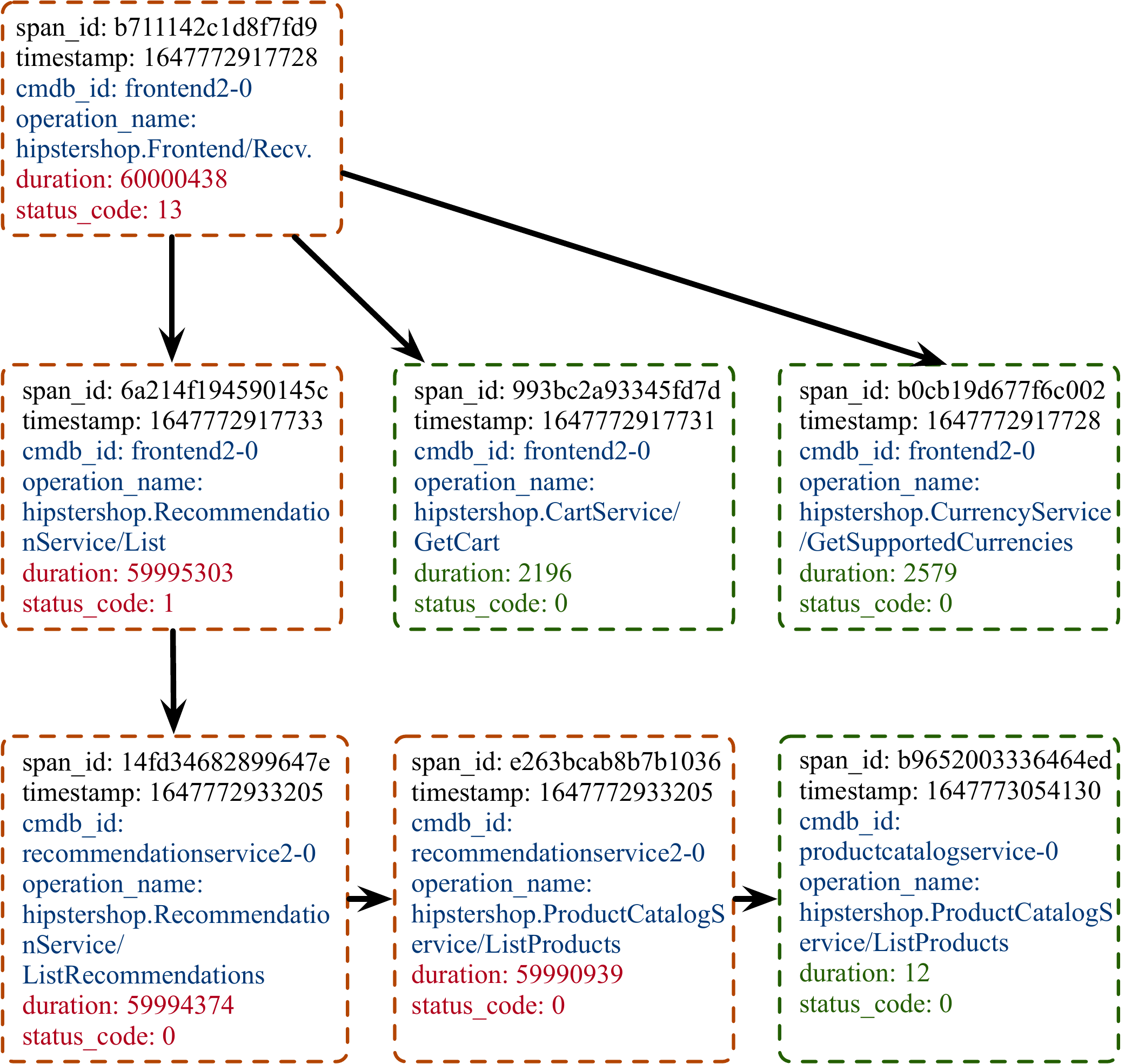}
	\vspace{-1.2em}
	\caption{Trace Graph Example-1}
	\label{fig: trace1}
	\vspace{-0.8em}
\end{figure}

For instance, we extracted a typical case from the real-world AIOPS 2022 dataset, as shown in Figure~\ref{fig: trace1}. In this case, a request with the cmdb\_id \texttt{frontend2-0} experienced severe timeouts, ultimately resulting in an error (status\_code = 13). As SREs, we first analyzed the trace data for this request and identified three primary operations: CartService/GetCart, RecommendationService/List, and CurrencyService/GetSupportedCurrencies. Notably, only the RecommendationService operation exhibited a timeout, indicating that it was likely responsible for the alert.

We then conducted a deeper investigation into the downstream services and found that the RecommendationService operation with cmdb\_id \texttt{recommendationservice2-0} also experienced a timeout. Further analysis revealed that the subsequent downstream operation, ProductCatalogService, under the same cmdb\_id, also timed out. However, a later operation on the component with cmdb\_id 'productcatalogservice-0' did not exhibit any issues.

\sloppy
This recursive investigation process effectively narrows down the potential root causes to the components with cmdb\_id \texttt{frontend2-0} and \texttt{recommendationservice2-0}, significantly reducing the search space for identifying the underlying performance issue.

\begin{center}
	\setlength{\fboxsep}{5pt} 
	\noindent\fcolorbox{black}{gray!10}{
		\begin{minipage}{0.93\linewidth} 
			\textbf{Summary.} By analyzing trace data, SREs narrow down the search space for the root cause by starting from the entry span of an alert trace and recursively examining downstream operations until the anomalous component is identified.
		\end{minipage}
	}
\end{center}

\subsection{Multi-Dimensional Expansion}

The previous analysis narrowed the search space by focusing on cmdb\_id, which represents individual pods. However, identifying anomalous cmdb\_ids alone is insufficient for pinpointing the true root causes, since failures often propagate across multiple components. In practice, SREs expand their investigation along two complementary dimensions: (i) \textbf{vertical expansion}, which explores deeper layers of the system hierarchy (service $\rightarrow$ service instance $\rightarrow$ pod $\rightarrow$ host) based on existing evidence, and (ii) \textbf{horizontal expansion}, which correlates alerts from the same anomaly window to uncover cross-service or multi-trace dependencies.  

\vspace{-0.8em}
\begin{figure}[htbp]
	\centering
	\includegraphics[width=1\linewidth]{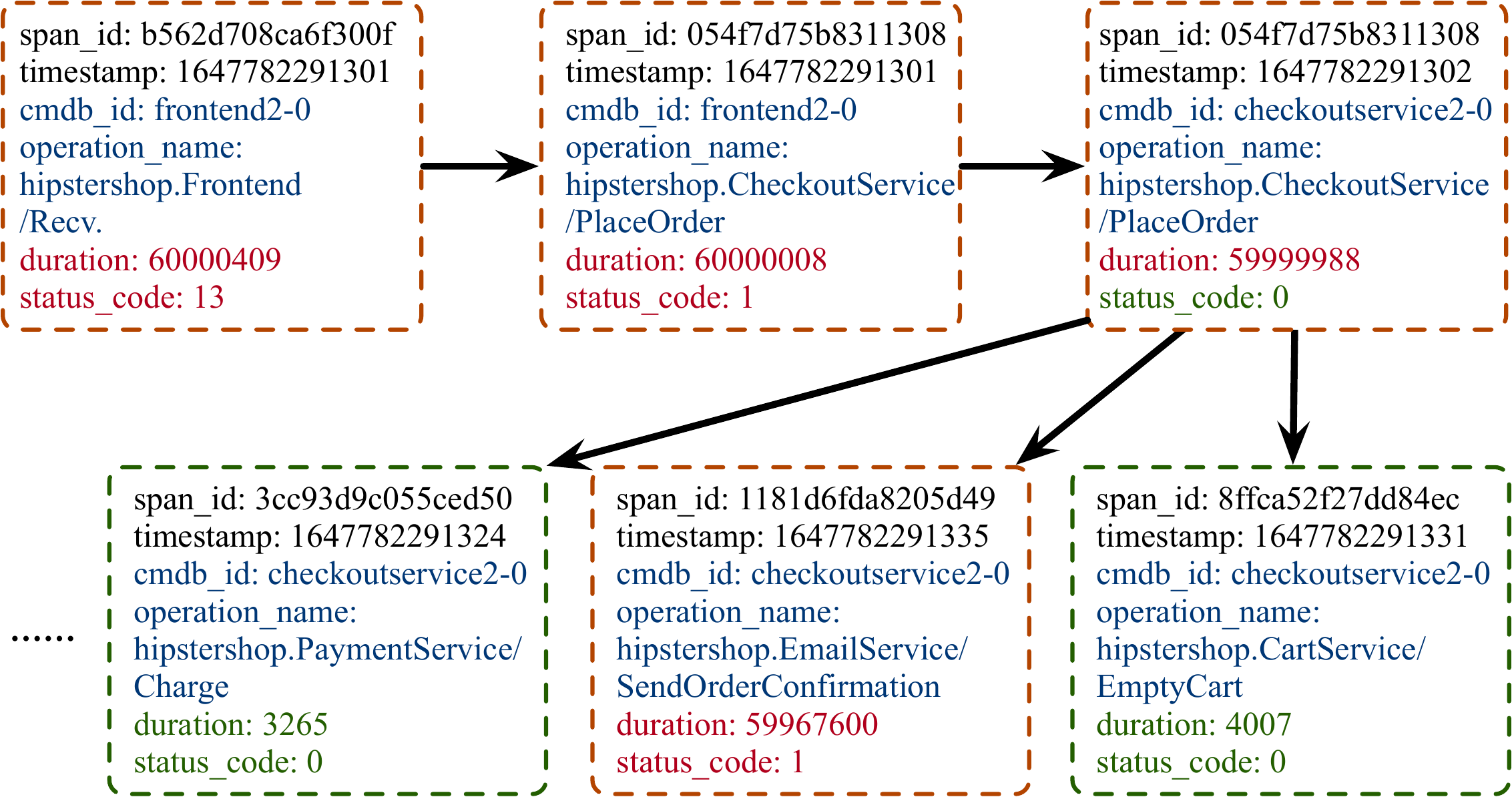}
	\vspace{-1.2em}
	\caption{Trace Graph Example-2}
	\label{fig: trace2}
	\vspace{-0.8em}
\end{figure}

\sloppy
For vertical expansion, operation\_name provides additional granularity by linking alerts to their corresponding services. From the recursive analysis above, for instance, both \texttt{recommendationservice} and \texttt{productcatalogservice} emerge as potential root causes. A more evident case is shown in Figure~\ref{fig: trace2}, where the anomaly is clearly tied to the \texttt{EmailService} operation under cmdb\_id \texttt{checkoutservice2-0}, indicating that further investigation into \texttt{EmailService} is warranted. Beyond the service layer, SREs also examine the underlying physical infrastructure. For example, both \texttt{recommendationservice2-0} and \texttt{checkoutservice2-0} are deployed on \texttt{node-5}, as observable in real-time from metrics data. This correlation suggests that the health of \texttt{node-5} itself must be assessed to determine whether it contributes to the observed anomalies.  

\vspace{-0.8em}
\begin{figure}[htbp]
	\centering
	\includegraphics[width=1\linewidth]{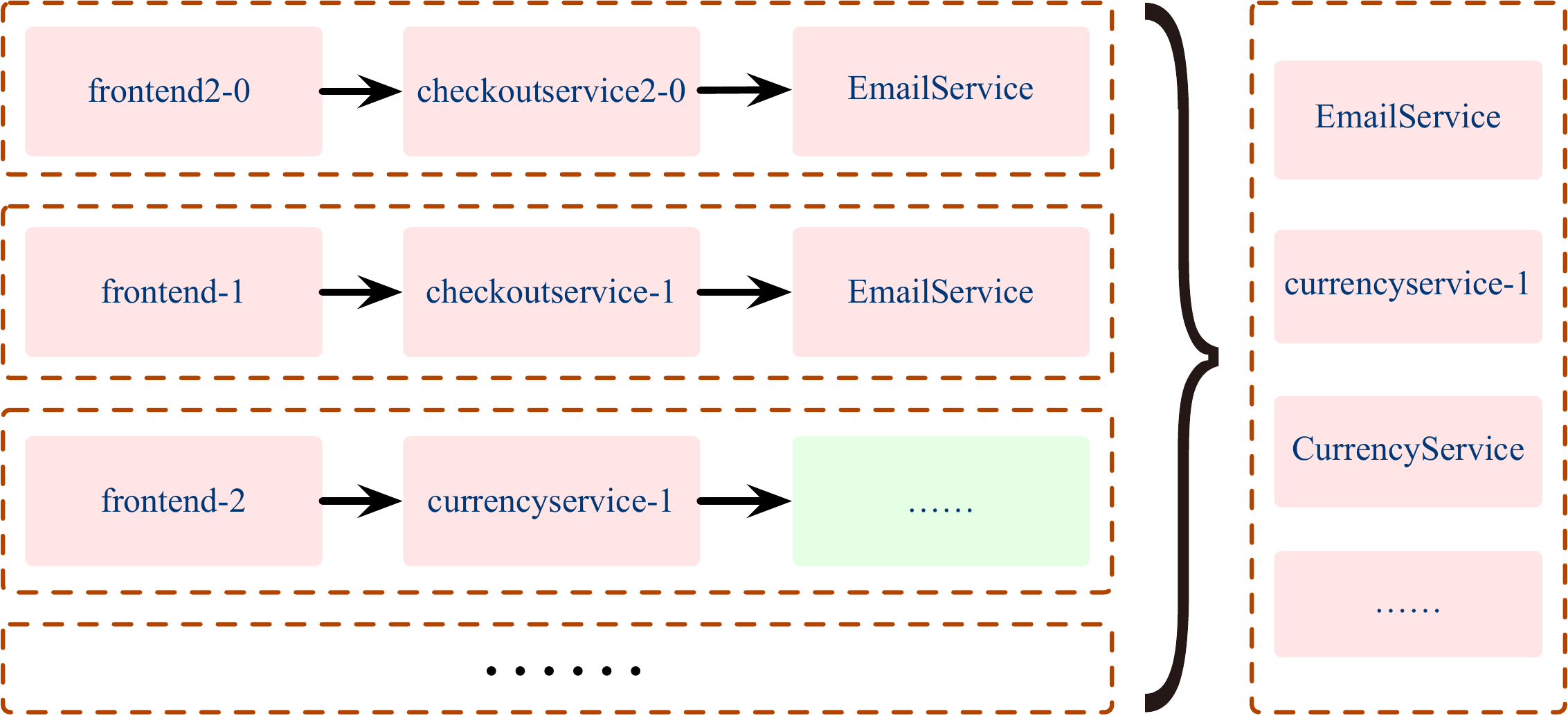}
	\vspace{-1.2em}
	\caption{Multi-Trace Graph Example}
	\label{fig: multi-trace}
	\vspace{-0.8em}
\end{figure}

\sloppy
For horizontal expansion, the presence of multiple alerts within the same anomaly window provides richer contextual evidence. As illustrated in Figure~\ref{fig: multi-trace}, several services eventually point to \texttt{EmailService} through different pods, while other alerts highlight errors in \texttt{currencyservice-1}. Such overlapping yet partially divergent signals necessitate comprehensive correlation across traces and alerts, enabling SREs to distinguish systemic issues from isolated failures and to refine the candidate set of root causes accordingly.

\begin{center}
	\setlength{\fboxsep}{5pt} 
	\noindent\fcolorbox{black}{gray!10}{
		\begin{minipage}{0.93\linewidth} 
			\textbf{Summary.} SREs expand root cause analysis along both vertical and horizontal dimensions—vertically by tracing anomalies from low-level instances to higher-level services and their underlying infrastructure, and horizontally by correlating alerts and traces within the same anomaly window—to more comprehensively uncover potential sources of failure.
		\end{minipage}
	}
\end{center}

\subsection{Cross-Modal Reasoning}

Through the analysis of recursiveness and multi-dimensional expansion, SREs have identified potential root causes across three dimensions: pod, service, and node. However, the exact root cause remains uncertain. At this stage, SREs often integrate other modalities of data, with metrics data being the most commonly used, to pinpoint the final root cause.

\vspace{-0.8em}
\begin{figure}[htbp]
	\centering
	\includegraphics[width=1\linewidth]{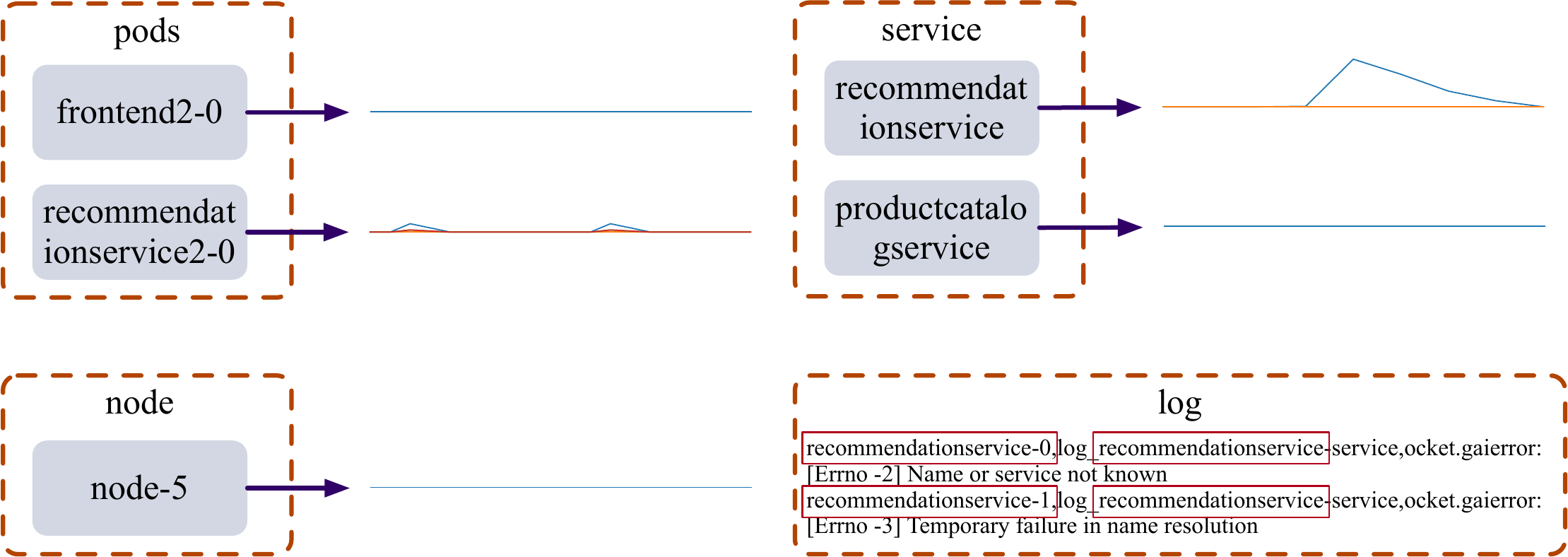}
	\vspace{-1.2em}
	\caption{Metrics \& Log Inspection of Trace Graph Example-1}
	\label{fig: metric-log}
	\vspace{-0.8em}
\end{figure}

For example, in Figure~\ref{fig: trace1}, the potentially problematic pods are \texttt{frontend2-0} and \texttt{recommendationservice2-0}, the potentially problematic services are \texttt{recommendationservice} and \texttt{productcatalogservice}, and the potentially problematic node is \texttt{node-5}. Analyzing the metrics data, as shown in Figure~\ref{fig: metric-log}, reveals that \texttt{frontend2-0}, \texttt{productcatalogservice}, and \texttt{node-5} exhibited no significant fluctuations. In contrast, both \texttt{recommendationservice2-0} and \texttt{recommendationservice} showed notable variations, with \texttt{recommendationservice} experiencing more pronounced fluctuations. The anomalies in \texttt{recommendationservice2-0} were propagated from \texttt{recommendationservice}. Additionally, the logs indicate that \texttt{recommendationservice-0} and \texttt{recommendationservice-1} both flagged \texttt{recommendationservice} as anomalous, indirectly ruling out an issue specific to any individual pod. Ultimately, the root cause is identified as \texttt{recommendationservice}, consistent with the ground truth—network packet corruption in the Kubernetes container.

\begin{center}
	\setlength{\fboxsep}{5pt} 
	\noindent\fcolorbox{black}{gray!10}{
		\begin{minipage}{0.93\linewidth} 
			\textbf{Summary.} SREs differentiate the actual root cause from other potential anomalies by analyzing other modalities of data, particularly metrics data, and identifying the component with the most significant or earliest fluctuation as the true root cause.
		\end{minipage}
	}
\end{center}

\section{AMER-RCL}

\begin{figure*}[htbp]
	\centering
	\includegraphics[width=1\linewidth]{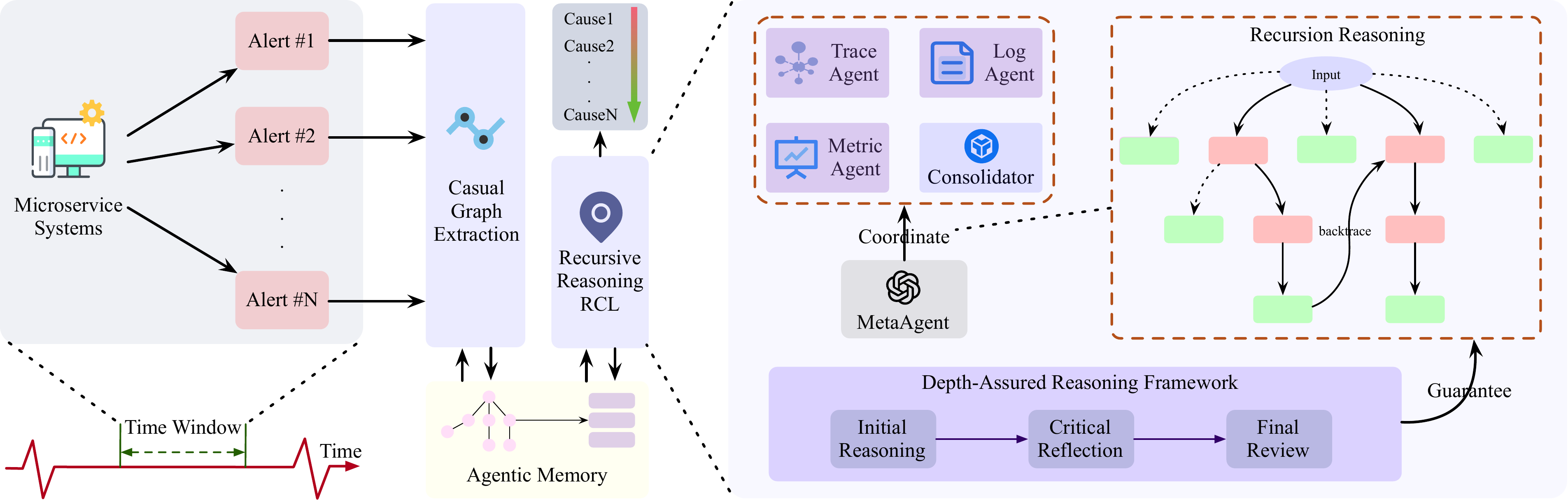}
	\caption{Architecture of AMER-RCL}
	\label{fig: architecture}
\end{figure*}

Our empirical study demonstrates how human SREs systematically localize root causes. Building on these findings, we introduce \textbf{AMER-RCL}, an \textbf{A}gentic \textbf{M}emory \textbf{E}nhanced \textbf{R}ecursive reasoning for \textbf{R}oot \textbf{C}ause \textbf{L}ocalization in microservices. Figure~\ref{fig: architecture} depicts the overall architecture of AMER-RCL.

AMER-RCL takes as input multiple alerts within a time window of a microservice system. Each alert is first transformed into a graph representation through a \textit{Causal Graph Extraction} process, which captures the anomalous request path and its contextual dependencies. These graphs serve as the basis for evaluating similarities and divergences among alerts. Leveraging \textit{Agentic Memory}, the system incrementally records both prior reasoning outcomes and intermediate deliberations from the \textit{Recursive Reasoning RCL} process, forming a persistent conversation history. This dual role enables the framework to skip redundant analyses for highly similar cases while guiding deeper exploration when significant divergences arise. As a result, the structured graph-to-conversation pipeline is tightly coupled with the \textit{Recursive Reasoning RCL} component, which conducts multi-agent reasoning across related alerts and ultimately generates a ranked list of candidate root causes.

The \textit{Recursive Reasoning RCL} module is driven by a \textit{MetaAgent}, which adopts a \textit{Recursive Reasoning} strategy to coordinate specialized agents such as the \textit{TraceAgent}, \textit{LogAgent}, and \textit{MetricAgent}, alongside a \textit{Consolidator} that integrates intermediate findings. To ensure both depth and reliability of this recursive process, the MetaAgent executes a \textit{Depth-Assured Reasoning Framework}: \textit{Initial Reasoning} for generating preliminary hypotheses, \textit{Critical Reflection} for scrutinizing and refining them, and \textit{Final Review} for synthesizing all evidence into the ranked root cause list.

\subsection{Agentic Memory}

Agentic Memory is designed as a structured, queryable repository that couples per-alert causal graphs (as keys) with their corresponding reasoning trajectories (as values). Conceptually, it is separable from the Recursive Reasoning RCL engine: causal graphs are extracted from raw alerts through deterministic data-processing methods, whereas the stored reasoning trajectories are leveraged at runtime to decide whether to reuse, resume from divergence, or re-initiate analysis when the engine is invoked on a new alert.

\subsubsection{Causal Graph Extraction}

For each input alert, we deterministically construct a causal graph, as illustrated in Equation~\ref{eq:graph}, where $\mathcal{V}$ is the set of nodes, $\mathcal{E}$ the set of directed edges, $\mathbf{A}_v$ the node attribute space, and $\mathbf{A}_e$ the edge attribute space.

\begin{equation}
	G = (\mathcal{V}, \mathcal{E}, \mathbf{A}_v, \mathbf{A}_e)
	\label{eq:graph}
\end{equation}

Each node $v \in \mathcal{V}$ is associated with an attribute vector $a_v = (\text{service\_id}, \text{pod\_id}, \text{op\_name}, \text{metric\_summary}, \text{log\_summary})$, where $\text{metric\_summary}_v = [\mu_m, \sigma_m]$ denotes the mean $\mu_m$ and standard deviation $\sigma_m$ of each metric over the alert window, and $\text{log\_summary}_v = [\text{count}_{l_1}, \dots, \text{count}_{l_k}]$ counts log entries of types $l_1,\dots,l_k$ observed within the same window.

Edges $(u,v) \in \mathcal{E}$ encode direct invocations or causal dependencies (e.g., service-to-service calls) and carry attributes $a_e = (\text{call\_latency}, \text{status\_code})$, where call\_latency is the observed duration of the call from $u$ to $v$ and status\_code is its response status.

\subsubsection{Memory Accumulation and Utilization}

After each execution of the Recursive Reasoning RCL, the corresponding causal graph and associated reasoning steps are stored in Agentic Memory as a memory entry, which can then be leveraged in subsequent RCL executions.

\textbf{Memory Accumulation.} Each entry is a tuple, as illustrated in Equation~\ref{eq:memory-entry}, where $f_i$ is a compact fingerprint of $G^{(i)}$, computed as a canonicalized graph hash to uniquely identify graph topology; $e_i$ is a dense embedding of $G^{(i)}$, obtained via a Graph2Vec~\cite{narayanan2017graph2vec} encoder, used for efficient nearest-neighbor retrieval; $G^{(i)}$ is the full causal graph for the alert; $\mathrm{meta}_i$ includes metadata such as timestamp and alert ID; and $\tau_i$ is the structured reasoning transcript, consisting of sequential reasoning steps, each explicitly linked to nodes in $G^{(i)}$.

\begin{equation}
	M_i = (f_i, e_i, G^{(i)}, \mathrm{meta}_i, \tau_i)
	\label{eq:memory-entry}
\end{equation}

\textbf{Memory Utilization.} Given a new alert graph $G^{\mathrm{new}}$, the system retrieves relevant entries from memory by first performing embedding-based nearest-neighbor search on $\{e_i\}$ and then computing exact similarity as Equation~\ref{eq:memory-utilization}, where $S_{\mathrm{struct}}$ measures structural similarity between graphs (e.g., via embedding cosine similarity or Weisfeiler-Lehman kernel), and $S_{\mathrm{attr}}$ measures node attribute agreement (e.g., normalized inverse distance over node-level metrics and logs).

\begin{equation}
	S(G^{\mathrm{new}}, G^{(i)}) = \alpha \cdot S_{\mathrm{struct}}(G^{\mathrm{new}}, G^{(i)}) + (1-\alpha) \cdot S_{\mathrm{attr}}(G^{\mathrm{new}}, G^{(i)})
	\label{eq:memory-utilization}
\end{equation}

Based on similarity thresholds, three scenarios arise: 
\begin{enumerate}
	\item \emph{High similarity}: if $S \ge \tau_{\mathrm{skip}}$, the stored reasoning $\tau_i$ can be reused directly.
	\item \emph{Partial similarity}: if $S \in [\tau_{\mathrm{partial}}, \tau_{\mathrm{skip}})$, divergence nodes
	\begin{equation}
		\mathcal{D} = \{v \in G^{\mathrm{new}} \mid \operatorname{dist}(a_v^{\mathrm{new}}, a_v^{(i)}) > \delta\}
		\label{eq:divergence}
	\end{equation}
	indicate points where reasoning should continue. Here, $\operatorname{dist}(\cdot,\cdot)$ is a normalized node attribute distance (e.g., Mahalanobis distance or z-score deviation) capturing differences in metrics, logs, or other diagnostic signals.
	\item \emph{Low similarity}: if $S < \tau_{\mathrm{partial}}$, reasoning is performed from scratch on $G^{\mathrm{new}}$.
\end{enumerate}

\subsection{Specialized Agents}

Specialized agents are responsible for performing focused analyses on different types of input data and intermediate information, according to their designated roles. Each specialized agent focuses on its designated role, processing traces, logs, metrics, or intermediate information to generate insights that contribute to the overall reasoning process. In this work, we primarily utilize four agents: the Trace Agent, Log Agent, Metric Agent, and Consolidator. The Trace, Log, and Metric Agents analyze their respective data sources and produce structured observations, while the Consolidator integrates these findings into a coherent summary for the MetaAgent to use in recursive reasoning.

\subsubsection{Trace Agent} 

Trace data is the most fundamental source for root cause localization, as it records the sequence of calls between various services. However, since each request generates a complete invocation path, the volume of trace data can be enormous, making it difficult for LLMs to process such extensive context. To address this, the trace agent is designed to filter and retrieve only the relevant subset of trace data.

\begin{equation}
	T(s) = \left\{ \langle t, s', svc, op, d, \sigma \rangle \; \middle| \; s' \in \mathcal{C}(s) \right\}
	\label{eq: trace-agent}
\end{equation}

Formally, given a span identifier $s$, the trace agent returns a set of child spans along with their associated metadata. We denote this function as Equation~\ref{eq: trace-agent}, where $t$ denotes the timestamp, $s'$ represents the child span identifier, $svc$ is the service name, $op$ is the operation name, $d$ is the duration, $\sigma$ is the status code, and $\mathcal{C}(s)$ is the set of all child spans for the given span $s$.

\subsubsection{Log Agent} 

Log data provide textual evidence of system events and component behaviors, which can be critical for diagnosing anomalies. However, the volume of logs is typically enormous, and only a subset is relevant to a particular incident. To address this, the Log Agent selectively retrieves logs that are most likely to be indicative of anomalies within a specified time window and target component.

Formally, given an input timestamp $t_0$ and a target key component $C$ (e.g., a pod or service), let $\mathcal{L}(C)$ denote the set of all logs associated with $C$ and its related entities. The Log Agent returns the subset of logs that fall within the interval $[t_0 - \delta, t_0 + \delta]$ and satisfy a relevance criterion, as formulated in Equation~\ref{eq: log-agent}.

\begin{equation}
	L(t_0, \delta, C) = 
	\left\{ 
	l \in \mathcal{L}(C) \;\middle|\; 
	t(l) \in [t_0 - \delta, t_0 + \delta] \wedge \phi(l) = 1 
	\right\}
	\label{eq: log-agent}
\end{equation}

Here, $t(l)$ denotes the timestamp of log entry $l$, and $\phi(l)$ is a binary function indicating whether $l$ is potentially relevant for anomaly analysis. In practice, $\phi(l)$ is determined by considering factors such as message type, error codes, and correlations with other alert or trace events. By filtering logs in this manner, the Log Agent efficiently reduces the context size while preserving the most informative entries for the recursive reasoning process.

\subsubsection{Metric Agent}

Metrics data represent the runtime state of system components. In a mature microservices environment, there are thousands of distinct metrics continuously recording values, often resulting in a data volume that exceeds that of trace data. Moreover, our empirical study reveals that during system anomalies, most metrics remain stable without significant fluctuations. Consequently, we have developed a specialized Metric Agent that selectively retrieves only those metrics exhibiting notable deviations within a predefined time window for LLM analysis.

\begin{equation}
	|m(t) - \mu_m| > n \times \sigma_m
	\label{eq: n-sigma}
\end{equation}

Formally, given an input timestamp $t_0$ and a target key component $C$ (e.g., a pod or service), let $\mathcal{M}(C)$ denote the set of metrics associated with $C$ and its related components (including the node corresponding to a pod). For each metric $m \in \mathcal{M}(C)$, let $\mu_m$ and $\sigma_m$ represent its historical mean and standard deviation, respectively. The Metric Agent examines the metric values over the interval $[t_0 - \delta, t_0 + \delta]$ and applies an $n$-sigma test: if there exists a time $t$ in this interval, as defined in Equation~\ref{eq: n-sigma}, an anomaly is detected. In such cases, the agent returns the corresponding fluctuation data over a period of length $\delta$, as formulated in Equation~\ref{eq: metric-agent}.

\begin{equation}
	Q(t_0, \delta, C) = 
	\left\{ 
	m(t) \;\middle|\; 
	\begin{array}{c}
		m \in \mathcal{M}(C),\; t \in [t_0 - \delta, t_0 + \delta], \\
		|m(t) - \mu_m| > n \times \sigma_m
	\end{array} 
	\right\}
	\label{eq: metric-agent}
\end{equation}

\subsubsection{Consolidator}

This agent serves as an integrative component that systematically aggregates intermediate inferences produced by the specialized agents during the recursive reasoning process. Rather than performing new analyses, it synthesizes the multi-step observations into a unified, structured representation that captures the essential evidential support for each hypothesized root cause. Formally, for each candidate root cause, the Consolidator produces a standardized output comprising two fields: \texttt{root\_cause}, denoting the identified source of the anomaly, and \texttt{reason}, encapsulating the aggregated rationale derived from the preceding agentic reasoning steps.

\subsection{MetaAgent}

The MetaAgent orchestrates the specialized agents under a recursive reasoning strategy, executing a depth-assured reasoning framework that guarantees both the depth and reliability of the process.

\subsubsection{Recursion Reasoning}

Recursive reasoning is inspired by our empirical study of how professional SREs localize root causes. Unlike traditional CoT reasoning, which proceeds in a linear fashion, recursive reasoning iteratively generates step-by-step instructions conditioned on prior outcomes. Each instruction activates subordinate agents to retrieve the necessary data for further analysis. When an analytical path is exhausted without yielding a plausible root cause, the process systematically backtracks to previously identified candidates that remain insufficiently explored, thereby ensuring a comprehensive and exhaustive search across the reasoning space.

\begin{algorithm}[htbp]
	\caption{Recursion Reasoning Algorithm}
	\label{alg:recursive-reasoning}
	\begin{algorithmic}[1]
		\REQUIRE Entry span $s_0$
		\ENSURE Ranked set of potential root causes $R$
		
		\STATE {\bf Function} \textsc{RecursiveRCL}($s$)
		\STATE $I \gets \text{MetaAgent.GenerateInstruction}(s)$
		\STATE $D_t \gets \text{TraceAgent}(I)$ \COMMENT{trace as primary exploration source}
		
		\IF{MetaAgent.Suspect($s, D_t$)}
		\STATE $D_l \gets \text{LogAgent}(I)$
		\STATE $D_m \gets \text{MetricAgent}(I)$
		\IF{MetaAgent.Confirm($s, D_l, D_m$)}
		\RETURN $\{s\}$ \COMMENT{confirmed as root cause}
		\ENDIF
		\ENDIF
		
		\STATE $R \gets \varnothing$
		\FOR{each child $c \in \text{MetaAgent.SuspiciousChildren}(s, D_t)$}
		\STATE $R \gets R \cup \text{RecursiveRCL}(c)$
		\ENDFOR
		\RETURN $R$
		\STATE {\bf End Function}
		\STATE
		\STATE $R \gets \text{RecursiveRCL}(s_0)$
		\RETURN $\text{Consolidator}(R)$
	\end{algorithmic}
\end{algorithm}

In this framework, recursive reasoning is primarily driven by trace data, with log and metrics data employed for cross-modal verification. As detailed in Algorithm~\ref{alg:recursive-reasoning}, analysis begins at the entry span of an alert trace. At each step, the \textit{MetaAgent} generates reasoning instructions for the current candidate span, and the \textit{TraceAgent} retrieves the corresponding trace information, serving as the primary basis for evaluating root cause potential.

Candidates flagged as suspicious are further examined using the \textit{LogAgent} and \textit{MetricAgent}, which provide contextual and system-level evidence to corroborate anomalies. Only spans confirmed across these modalities are added to the potential root cause set. For unconfirmed spans, the \textit{MetaAgent} recursively identifies and evaluates suspicious child spans. This process continues until no viable candidates remain, with backtracing employed to ensure comprehensive coverage of unexplored paths. The \textit{Consolidator} then synthesizes all intermediate results into a structured, ranked list of root causes.

\subsubsection{Depth-Assured Reasoning Framework}

While the recursive reasoning algorithm (Algorithm~\ref{alg:recursive-reasoning}) provides a principled mechanism for root cause localization, we observe that LLMs often terminate the recursion prematurely, overlooking deeper anomalies. To address this issue, we introduce a \textit{depth-assured reasoning framework} that explicitly enforces sufficient exploration depth during the execution of \texttt{RecursiveRCL}. This framework is structured into three complementary stages---\textit{Initial Reasoning}, \textit{Critical Reflection}, and \textit{Final Review}---each of which regulates the recursive reasoning process to ensure both depth and reliability.

\textbf{Initial Reasoning.} At the beginning of \texttt{RecursiveRCL}, the MetaAgent generates reasoning instructions from the entry span $s_0$, and the TraceAgent conducts broad exploration of the trace tree. To accelerate the early expansion of candidate spans, the Metrics and Log Agents are deliberately withheld in this stage. The initial recursion thus yields a preliminary reasoning trajectory, as illustrated in Equation~\ref{eq:initial-reasoning}, where $A$ denotes the set of specialized agents. This step provides a coarse-grained search frontier for subsequent refinement.

\begin{equation}
	\Gamma_0 = \texttt{RecursiveRCL}\bigl(s_0 \,\big|\, A \setminus \{A_{\text{metric}}, A_{\text{log}}\}\bigr)
	\label{eq:initial-reasoning}
\end{equation}

\textbf{Critical Reflection.} As \texttt{RecursiveRCL} progresses, the model may stop once a plausible candidate is found, leading to shallow reasoning. To counteract this tendency, the critical reflection stage forces the continuation of recursion on all spans in $\Gamma_0$ that remain suspicious. In this stage, the Metrics and Log Agents are incorporated to confirm or refute anomalies, thereby deepening the recursive exploration. Formally, the refined reasoning trajectory is obtained by Equation~\ref{eq:critical-reflection}, ensuring that deeper layers of the trace tree are systematically inspected.

\begin{equation}
	\Gamma_1 = \bigcup_{s \in \Gamma_0} \texttt{RecursiveRCL}(s \,|\, A)
	\label{eq:critical-reflection}
\end{equation}

\textbf{Final Review.} Even after recursive refinement, LLMs may exhibit recency bias, overemphasizing spans encountered in later recursion. To mitigate this, the final review stage consolidates the outputs of all recursive calls across both $\Gamma_0$ and $\Gamma_1$. This stage invokes only the Consolidator, which synthesizes the intermediate results into the final ranked root cause set, as illustrated in Equation~\ref{eq:final-review}.

\begin{equation}
	R_f = A_{\text{consolidator}}(\Gamma_0 \cup \Gamma_1)
	\label{eq:final-review}
\end{equation}

By reviewing the entire recursive reasoning trajectory, the framework guarantees that no candidate root cause is prematurely discarded.

\section{Evaluation}

To evaluate AMER-RCL, we conduct a series of experimental studies to investigate the following research questions:

\begin{itemize}[leftmargin=*]
	\item \textbf{EV-RQ1:} How accurate is AMER-RCL in root cause localization when compared with baseline approaches? 
	\item \textbf{EV-RQ2:} How efficient is AMER-RCL in terms of inference latency when compared with LLM-based approaches?
	\item \textbf{EV-RQ3:} How does the choice of different LLM backbones affect the accuracy of AMER-RCL? 
	\item \textbf{EV-RQ4:} What is the contribution of each component of AMER-RCL to its overall accuracy? 
\end{itemize}

\subsection{Experimental Setup}

\subsubsection{Dataset} 

We evaluate our approach on three representative datasets: the AIOPS 2022 companion~\cite{aiops2022championship} dataset, the TrainTicket dataset~\cite{zhou2018fault, zhou2018benchmarking}, and the FAMOS-Mall~\cite{duan2025famos} dataset.  

\textbf{AIOPS 2022} is a large-scale real-world dataset collected from a mature, actively running microservices-based e-commerce platform. It contains rich operational data and diverse anomaly cases, making it a widely used benchmark for anomaly detection and diagnosis.  

\textbf{TrainTicket} is an open-source microservice benchmark system that simulates typical train ticket booking services and has been extensively adopted in prior research on failure diagnosis. We deploy TrainTicket on a Kubernetes cluster consisting of six virtual machines, each equipped with an 8-core Intel Xeon 2.60GHz CPU and 24 GB RAM, running CentOS 8. In total, the deployment comprises 90 service instances. To evaluate system robustness, we inject failures using ChaosMesh, covering ten distinct fault types.  

\textbf{FAMOS-Mall} is an e-commerce backend microservice system that provides core services such as user management, shopping cart, and product management. Unlike the synthetic benchmark, FAMOS-Mall is deployed in a real production environment and serves as a demonstration system for Alibaba Cloud’s Application Real-Time Monitoring Service (ARMS), thereby reflecting realistic operational conditions.  

\subsubsection{Baseline Approaches}

We primarily compare AMER-RCL with three representative LLM-based root cause localization approaches. mABC~\cite{zhang2024mabc} adopts a multi-agent, blockchain-inspired collaboration framework, where each alert is analyzed individually by multiple LLM-based agents, and the final root cause is determined via a voting mechanism. RCAgent~\cite{wang2024rcagent} employs a ReAct-based reasoning-and-acting paradigm, processing multiple alerts jointly: a controller agent executes a thought–action–observation loop, invoking various tools to analyze alerts collectively. CoT-based Approach~\cite{wei2022chain} implements a naive chain-of-thought reasoning strategy, handling alerts individually and generating conclusions sequentially.

To provide a more comprehensive comparison, we also include several non-LLM-based approaches. (1) \textbf{Trace-based methods:} CRISP~\cite{zhang2022crisp} represents traces as critical paths and applies lightweight heuristics to identify root causes. TraceContrast~\cite{zhang2024trace} employs sequence representations, contrastive sequential pattern mining, and spectrum analysis to localize multi-dimensional root causes. TraceRCA~\cite{li2021practical} uses frequent itemset mining and spectrum analysis to locate the root cause instances. MicroRank~\cite{yu2021microrank} constructs a trace coverage tree to capture dependencies between requests and service instances, leveraging the PageRank algorithm to rank potential root causes.  (2) \textbf{Metrics-based methods:} RUN~\cite{lin2024root} applies time series forecasting for neural Granger causal discovery and combines it with a personalized PageRank algorithm to recommend top-$k$ root causes. Microscope~\cite{lin2018microscope} builds causality graphs and employs a depth-first search strategy to detect anomalies in front-end services. 

\begin{table*}[tbp]
	\setlength{\tabcolsep}{2.4pt}
	\centering
	\caption{Root Cause Localization Accuracy Compared with SOTA Methods (R$k$ denotes Recall@$k$).}
	\label{tab: accuracy}
	\vspace{-1.2em}
	\begin{tabular}{c|c|ccccc|ccccc|ccccc}
		\toprule
		\multirow{2}{*}{\textbf{Paradigm}} & \multirow{2}{*}{\textbf{Method}} & \multicolumn{5}{c|}{AIOPS 2022} & \multicolumn{5}{c|}{Train-Ticket} & \multicolumn{5}{c}{FAMOS-Mall} \\
		\\[-2.5ex]
		\cline{3-17}
		\\[-2ex]
		~ & ~ & \textit{\textbf{R1}} & \textit{\textbf{R3}} & \textit{\textbf{R5}} & \textit{\textbf{R10}} & \textit{\textbf{MRR}} & \textit{\textbf{R1}} & \textit{\textbf{R3}} & \textit{\textbf{R5}} & \textit{\textbf{R10}} & \textit{\textbf{MRR}} & \textit{\textbf{R1}} & \textit{\textbf{R3}} & \textit{\textbf{R5}} & \textit{\textbf{R10}} & \textit{\textbf{MRR}} \\
		\midrule
		\multirow{6}{*}{\textit{non-LLM-based}} & CRISP & 3.92 & 29.79 & 48.88 & 51.01 & 18.68 & 31.45 & 41.77 & 48.19 & 55.26 & 38.53 & 24.81 & 44.48 & 50.83 & 60.52 & 35.34 \\
		~ & TraceConstract & 24.06 & 36.51 & 37.69 & 40.81 & 30.29 & 48.37 & 55.33 & 63.78 & 78.21 & 54.35 & 45.37 & 72.83 & 82.84 & 89.34 & 58.71 \\
		~ & TraceRCA & 36.43 & 43.13 & 44.56 & 51.88 & 42.19 & 42.73 & 46.98 & 47.17 & 49.32 & 45.34 & 30.13 & 51.08 & 57.96 & 69.18 & 41.17 \\
		~ & MicroRank & 8.45 & 24.57 & 37.24 & 42.89 & 18.42 & 24.15 & 42.10 & 50.16 & 62.33 & 34.73 & 22.99 & 43.94 & 50.05 & 59.72 & 33.69 \\
        ~ & RUN & 4.69 & 11.23 & 11.88 & 25.81 & 10.44 & 40.03 & 47.19 & 51.31 & 56.77 & 48.38 & 34.33 & 56.07 & 64.21 & 75.34 & 46.70 \\
		~ & MicroScope & 9.23 & 28.81 & 33.37 & 51.25 & 20.39 & 32.37 & 40.18 & 43.26 & 48.70 & 36.88 & 26.49 & 47.84 & 56.30 & 65.08 & 37.91 \\
		\midrule
		\multirow{3}{*}{\textit{LLM-based}} & CoT & 8.96 & 15.89 & 34.72 & 48.33 & 17.56 & 22.12 & 39.75 & 48.66 & 61.19 & 32.14 & 31.16 & 50.94 & 58.13 & 66.16 & 42.33 \\
		~ & RCAgent & 13.15 & 23.11 & 38.56 & 54.87 & 22.53 & 38.51 & 53.14 & 60.47 & 69.98 & 47.39 & \underline{46.86} & \underline{75.17} & \underline{83.10} & \underline{91.18} & \underline{61.05} \\
		~ & mABC & \underline{51.68} & \underline{60.33} & \underline{65.33} & \underline{78.46} & \underline{58.05} & \underline{57.53} & \underline{74.67} & \underline{83.15} & \underline{90.02} & \underline{67.26} & 45.39 & 74.11 & 82.53 & 89.93 & 59.96 \\
		\midrule
		\multicolumn{2}{c|}{AMER-RCL (\textit{ours})} & \textbf{68.67} & \textbf{81.50} & \textbf{90.36} & \textbf{96.71} & \textbf{76.30} & \textbf{72.97} & \textbf{89.15} & \textbf{93.47} & \textbf{96.52} & \textbf{81.93} & \textbf{66.55} & \textbf{84.38} & \textbf{90.06} & \textbf{95.58} & \textbf{75.39} \\
		\bottomrule
	\end{tabular}
\end{table*}

\subsubsection{Evaluation Metrics and Settings}

We evaluate root cause localization accuracy using top-$k$ recall (Recall@$k$) and mean reciprocal rank (MRR), following prior work~\cite{zhang2024trace, yu2021microrank, lin2018microscope}. Unless otherwise specified, we use Claude-3.5 Sonnet as the LLM backbone, set the $n$-sigma threshold in the Metric Agent to $n=3$, and set the anomaly window size to 60 seconds.

\subsection{Accuracy}

We compare AMER-RCL against both non-LLM-based and LLM-based methods on three benchmark datasets. As shown in Table~\ref{tab: accuracy}, AMER-RCL consistently outperforms all SOTA baselines. On the AIOPS 2022 dataset, AMER-RCL achieves an MRR of 76.30\%, surpassing the strongest baseline mABC by 18.25\%, and the best non-LLM-based model TraceRCA by 34.11\%. On the Train-Ticket dataset, AMER-RCL attains an MRR of 81.93\%, exceeding mABC by 14.67\% and TraceConstract by 27.58\%. On the FAMOS-Mall dataset, AMER-RCL reaches an MRR of 75.39\%, outperforming mABC by 14.34\% and TraceConstract by 16.68\%. Overall, averaged across all datasets, AMER-RCL improves upon the second-best method (mABC) by 15.75\%.

Notably, mABC outperforms RCAgent on the AIOPS 2022 and Train-Ticket datasets, whereas RCAgent fares better on FAMOS-Mall. We attribute this divergence to differences in alert characteristics and how each method ingests evidence. FAMOS-Mall alerts are relatively homogeneous, so RCAgent’s batch-style, joint analysis can exploit early, consistent corroboration across alerts and reach confident conclusions. By contrast, alerts in AIOPS and Train-Ticket are more heterogeneous; when many disparate alerts are fed together, batch input can exceed pragmatic context and attention limits, causing the model to prioritize a subset of evidence and prematurely terminate deeper inquiry. mABC’s strategy of analyzing alerts independently and aggregating results via voting is more robust to such heterogeneity, but it is computationally redundant and can produce noisy or inconsistent votes when local analyses disagree. AMER-RCL avoids these pitfalls by (i) reusing prior reasoning for highly similar alerts to avoid redundant work, and (ii) launching targeted, depth-assured recursive analysis at divergence points to fully explore heterogeneous cases. This hybrid behavior explains why AMER-RCL maintains superior performance across both homogeneous and heterogeneous datasets.

\subsection{Efficiency}

We further evaluate AMER-RCL in terms of inference efficiency against representative LLM-based methods. As shown in Table~\ref{tab: efficiency}, AMER-RCL achieves substantial speedups: it is on average 3.53$\times$ faster than RCAgent, 31.05$\times$ faster than mABC, and 14.61$\times$ faster than CoT, demonstrating its practical efficiency.

\vspace{-0.8em}
\begin{table}[htb]
	\setlength{\tabcolsep}{4.5pt}
	\centering
	\caption{Inference Speed Comparison (seconds/query)}
	\label{tab: efficiency}
	\vspace{-1.2em}
	\begin{tabular}{c|ccc}
		\toprule
		Method & AIOPS 2022 & Train-Ticket & FAMOS-Mall \\
		\midrule
		CoT & 1517.34 & 543.58 & 1401.51 \\
		RCAgent & \underline{322.16} & \underline{305.84} & \underline{378.23} \\
		mABC & 2731.89 & 1490.15 & 2890.12 \\
		AMER-RCL (\textit{ours}) & \textbf{111.45} & \textbf{53.86} & \textbf{56.59} \\
		\bottomrule
	\end{tabular}
	\vspace{-0.8em}
\end{table}

RCAgent is relatively fast because it processes multiple alerts in a single batch, avoiding repeated initialization, whereas mABC and CoT analyze alerts individually, with CoT being slightly faster due to shallower reasoning. Notably, AMER-RCL outperforms RCAgent by leveraging Agentic Memory to reuse prior reasoning outcomes, thereby skipping redundant computations. The speedup is particularly pronounced on the FAMOS-Mall dataset, where alerts are largely homogeneous: AMER-RCL can converge by analyzing only representative alerts, while RCAgent, despite batch processing, still performs some redundant reasoning across similar alerts. These results highlight AMER-RCL’s ability to combine recursive in-depth reasoning with efficient cross-alert reuse.

\subsection{LLM Backbone Impact}

Next, we address EV-RQ3. Using the AIOPS 2022 dataset—which is further divided into six subsets based on recorded cloudbed and date, labeled as $\mathbf{A}$, $\mathbf{B}$, $\mathbf{\Gamma}$, $\mathbf{\Delta}$, $\mathbf{E}$, and $\mathbf{Z}$—we evaluate the performance of AMER-RCL across different LLM backbones. In addition to Claude-3.5-Sonnet, we conduct experiments with four state-of-the-art LLMs: DeepSeek-R1-Qwen, a Qwen2.5-32B model fine-tuned on distilled data released by DeepSeek; Qwen-2.5-Max and Qwen-2.5-Plus, both closed-source models accessed via API; and Llama-3.1-70B, the original open-source model released by Meta.

\vspace{-0.8em}
\begin{table}[htb]
	\setlength{\tabcolsep}{3pt}
	\centering
	\caption{AMER-RCL with different LLMs (MRR)}
	\label{tab: llm-impact}
	\vspace{-1.2em}
	\begin{tabular}{c|cccccc}
		\toprule
		Model & $\mathbf{A}$ & $\mathbf{B}$ & $\mathbf{\Gamma}$ & $\mathbf{\Delta}$ & $\mathbf{E}$ & $\mathbf{Z}$ \\
		\midrule
		Claude-3.5-sonnet & \textbf{73.89} & \textbf{81.23} & \textbf{92.87} & \textbf{66.81} & \textbf{68.05} & \textbf{74.92} \\
		DeepSeek-R1-qwen & 35.21 & 60.98 & \underline{70.12} & \underline{71.63} & \underline{47.21} & \underline{63.78} \\
		Qwen-2.5-max & 12.21 & 20.12 & 32.14 & 12.21 & 32.14 & 12.71 \\
		Qwen-2.5-plus & \underline{41.83} & \underline{72.87} & 54.91 & 51.62 & 37.58 & 54.19 \\
		Llama-3.1-70B & 6.81 & 21.32 & 16.23 & 16.91 & 5.86 & 7.65 \\
		\bottomrule
	\end{tabular}
\vspace{-0.8em}
\end{table}

As shown in Table~\ref{tab: llm-impact}, the performance of AMER-RCL is closely tied to the reasoning capabilities of the underlying LLM backbone. In our experiments, Claude-3.5-Sonnet consistently achieved the best results, followed by DeepSeek-R1-Qwen. However, due to its relatively smaller size (32B), DeepSeek-R1-Qwen underperformed on datasets A and B compared to Qwen-2.5-Plus. Overall, AMER-RCL with Claude-3.5-Sonnet outperformed the second-best model by an average of 18.14\%. These findings highlight that AMER-RCL effectively leverages the semantic understanding and logical reasoning capabilities of LLMs, achieving better results when paired with more powerful inference engines.

\subsection{Ablation Study}

To address EV-RQ4, we recorded the complete reasoning path of AMER-RCL and evaluated the contributions of its key components: the Depth-Assured Reasoning Framework—comprising Initial Reasoning (IR), Critical Reflection (CR), and Final Review (FR)—together with Agentic Memory (Mem).

\vspace{-0.8em}
\begin{table}[htb]
	\setlength{\tabcolsep}{4.9pt}
	\centering
	\caption{Ablation Experiment (MRR)}
	\label{tab: ablation}
	\vspace{-1.2em}
	\begin{tabular}{c|cccccc}
		\toprule
		Method & $\mathbf{A}$ & $\mathbf{B}$ & $\mathbf{\Gamma}$ & $\mathbf{\Delta}$ & $\mathbf{E}$ & $\mathbf{Z}$ \\
		\midrule
		\textbf{AMER-RCL} & \textbf{73.89} & \textbf{81.23} & \textbf{92.87} & \textbf{66.81} & \textbf{68.05} & \textbf{74.92} \\
		\midrule
		w/o Mem & 71.13 & 78.57 & 90.24 & 64.34 & 65.42 & 72.41 \\
		w/o FR & 70.45 & 72.65 & 84.15 & 64.13 & 61.23 & 71.77 \\
		w/o CR \& FR & 35.93 & 71.75 & 68.14 & 61.31 & 41.41 & 66.58 \\
		\bottomrule
	\end{tabular}
\vspace{-0.8em}
\end{table}

As shown in Table~\ref{tab: ablation}, Initial Reasoning alone achieves an average MRR of 57.52\% across the six subsets, serving as the baseline. Incorporating Critical Reflection substantially improves localization, for example raising MRR on subset $\mathbf{A}$ from 35.93\% to 70.45\% and on subset $\mathbf{E}$ from 41.41\% to 61.23\%, demonstrating the effectiveness of enforced deep reasoning. Final Review further strengthens performance, yielding additional gains such as 72.65\% $\rightarrow$ 78.57\% on subset $\mathbf{B}$ and 84.15\% $\rightarrow$ 90.24\% on subset $\mathbf{\Gamma}$. Finally, Agentic Memory provides consistent improvements by leveraging prior reasoning outcomes, e.g., 71.13\% $\rightarrow$ 73.89\% on subset $\mathbf{A}$ and 72.41\% $\rightarrow$ 74.92\% on subset $\mathbf{Z}$, underscoring its role in reducing redundant reasoning and consolidating knowledge across multiple alerts.

Overall, incorporating Critical Reflection yielded an average improvement of 13.21\%, Final Review contributed a further 2.96\% boost, and Agentic Memory added another 2.61\%. These findings highlight the effectiveness and soundness of AMER-RCL’s design.

\subsection{Threats to Validity}

Despite the promising results demonstrated by AMER-RCL, several limitations should be acknowledged. First, the implementation and configuration of baseline approaches: CRISP, MicroRank, and mABC have publicly available source code, which we directly utilized. For other state-of-the-art methods, we implemented them based on the descriptions in their respective papers. After implementation, we fine-tuned the results and selected the best configurations for all baselines through experimentation. Second, although our empirical study benefits from insights provided by multiple professional SREs, the inherent subjectivity in their diagnostic processes may introduce bias.

\section{Related Work}

\subsection{Root Cause Localization}

Root cause localization aims to identify the services, components, or operations responsible for system anomalies. Existing methods typically leverage either system metrics or distributed traces, providing complementary perspectives: metrics highlight abnormal resource usage or performance deviations, while traces reveal detailed service interactions and request flows.

Metrics-based approaches analyze KPIs such as response times, throughput, and resource utilization. Microscope~\cite{lin2018microscope} constructs causality graphs and employs depth-first search to detect front-end anomalies. CIRCA~\cite{li2022causal} builds a causal Bayesian network using regression-based hypothesis testing and descendant adjustment to infer problematic components. RUN~\cite{lin2024root} integrates time series forecasting for neural Granger causal discovery with a personalized PageRank algorithm to efficiently recommend the top-k root causes.

Trace-based approaches track execution paths and inter-component interactions for fine-grained localization. MicroRank~\cite{yu2021microrank} builds a trace coverage tree and applies PageRank to score potential root causes. TraceRank~\cite{yu2023tracerank} combines spectrum analysis with PageRank-based random walks to pinpoint abnormal services. CRISP~\cite{zhang2022crisp} performs critical path analysis to drill down on latency issues, while TraceConstruct~\cite{zhang2024trace} leverages sequence representations, contrast sequential pattern mining, and spectrum analysis to localize multi-dimensional root causes efficiently.

\subsection{LLM-based Failure Management}

Large language models, with their advanced semantic understanding and logical reasoning capabilities, have significantly improved the field of failure management~\cite{zhang2025survey} and are increasingly becoming a focal point of research. Numerous LLM-based approaches have been proposed to address various aspects of failure management, including anomaly detection, failure diagnosis, and automated mitigation~\cite{rasul2023lag, liu2024timer, das2024decoder, shi2023shellgpt, liu2024anomalyllm, liu2024unitime, guo2023owl, liu2024loglm, chen2024automatic, jiang2024xpert, zhang2024lm, hamadanian2023holistic, pan2024raglog, zhang2024lograg, zhang2025xraglog, zhai2025agentevolver, zhang2025logdb, liu2025ora, zhang2025surveyparallel, pan2025omni, pan2025d}.

Some studies have developed foundation models specifically for failure management. For example, Lag-Llama~\cite{rasul2023lag}, Timer~\cite{liu2024timer}, and TimesFM~\cite{das2024decoder} pretrain foundation models for metrics-based anomaly detection. Similarly, ShellGPT~\cite{shi2023shellgpt} trains a model capable of automatically generating shell scripts for automated mitigation. Other approaches adopt fine-tuning strategies to tailor LLMs for failure management tasks. For instance, AnomalyLLM~\cite{liu2024anomalyllm} and UniTime~\cite{liu2024unitime} employ full fine-tuning for anomaly detection, while OWL~\cite{guo2023owl} and LogLM~\cite{liu2024loglm} leverage parameter-efficient fine-tuning techniques to build log analysis models.

Since these training-based approaches require significant computational resources, an increasing number of methods rely on prompt-based techniques. For example, RCACopilot~\cite{chen2024automatic} and Xpert~\cite{jiang2024xpert} utilize in-context learning to structure diagnostic processes, ensuring accurate root cause analysis. LM-PACE~\cite{zhang2024lm} applies CoT reasoning to enhance GPT-4’s ability to analyze incident reports, while Hamadanian et al.~\cite{hamadanian2023holistic} extend this approach to generate mitigation solutions from incident reports. Additionally, RAGLog~\cite{pan2024raglog} and LogRAG~\cite{zhang2024lograg} use retrieval-augmented generation (RAG) to enhance log-based anomaly detection through historical log retrieval.

\section{Conclusion}

In this paper, we introduce AMER-RCL, an agentic memory enhanced recursive reasoning framework for root cause localization in microservices. The design of AMER-RCL is motivated by our empirical study of how SREs perform root cause analysis. It employs a Recursive Reasoning RCL engine to recursively refine in-depth candidate causes for each alert and leverages Agentic Memory to incrementally accumulate and reuse reasoning from previous alerts within the same time window, thereby reducing redundant exploration and lowering inference latency. Experimental results demonstrate that AMER-RCL consistently outperforms state-of-the-art methods in both localization accuracy and inference efficiency.

In future work, we will further explore how to achieve more accurate and efficient root cause localization using smaller-scale models. Additionally, we are considering extending our approach to cover the entire failure management process.

\begin{acks}
This work is supported by Key RD Project of Guangdong Province, China (No.2020B010164003).
\end{acks}

\bibliographystyle{ACM-Reference-Format}
\balance
\bibliography{sample-base}

\end{document}